\documentclass{PoS}

\usepackage{amsfonts}
\usepackage{amsmath}
\usepackage{subfigure} 

\graphicspath{{./figs/}}

\title{
   Finite Volume Errors in $B_K$  
}

\ShortTitle{
   Finite Volume Errors in $B_K$
}

\author{\speaker{Jangho Kim}, 
  Hyung-Jin Kim, Weonjong Lee \\
  Lattice Gauge Theory Research Center, CTP, and FPRD, \\
  Department of Physics and Astronomy,
  Seoul National University, Seoul, 151-747, South Korea \\
  E-mail: \email{wlee@snu.ac.kr}}

\author{Chulwoo Jung \\
  Physics Department, Brookhaven National Laboratory,
  Upton, NY11973, USA \\
  E-mail: \email{chulwoo@bnl.gov}}

\author{Stephen R. Sharpe\\
  Physics Department, University of Washington, 
  Seattle, WA 98195-1560, USA \\
  E-mail: \email{sharpe@phys.washington.edu}}

\author{SWME Collaboration}

\abstract{
  We discuss finite volume errors in our calculations of
  $B_K$ using improved staggered fermions on the MILC asqtad
  lattices.
  Using GPUs, we are now able to extrapolate using
  next-to-leading order (NLO) staggered SU(2) chiral perturbation theory 
  including  the finite volume corrections arising from pion loops.
  We find that the impact of FV fitting is very small,
  giving a $0.5\%$ shift in the continuum limit.
}

\FullConference{
  The XXIX International Symposium on Lattice Field Theory - Lattice 2011\\
    July 10-16, 2011\\
    Squaw Valley, Lake Tahoe, California
}

\begin{document}

\section{Introduction \label{sec:intr}} 
The dominant error in our calculation of $B_{K}$ 
using improved staggered quarks~\cite{wlee-10-3}
comes from our use
of a truncated matching factor, but a significant
subdominant error is that from extrapolating from
finite to infinite volume. 
In our earlier work we estimated this error by comparing
the results on two volumes.
Here we describe an alternative estimate using 
next-to-leading order (NLO) chiral perturbation theory (ChPT).
Specifically, we replace the pion loop integrals with their
finite volume form, perform the chiral fit, and then use the
fit parameters to determine the result in infinite volume.
This method is fairly standard in chiral fits, but, given our
large data set, has been too expensive to implement at the
desired accuracy until recently.
Using GPUs we can now incorporate the finite volume (FV) corrections
into the fitting routines using $SU(2)$ staggered ChPT.
First results were presented in Ref.~\cite{wlee-11-1},
and here we present an update.
\section{Finite Volume Effects in SU(2) staggered chiral perturbation
  theory\label{sec:su2}}
Finite volume corrections enter at NLO in SU(2) ChPT only through 
the chiral logarithms arising from loops of pions composed
of valence $\bar d$ and $d$ quarks.
The standard chiral logarithmic functions that enter are 
\begin{eqnarray}
\ell(X) &=& X \left[\log(X/\mu_{\rm DR}^2) 
+\delta^{\rm FV}_1(X) \right]\,,
\label{eq:app:l}
\\
\tilde\ell(X) &=& - \frac{d\ell(X)}{dX} 
=
-\log(X/\mu_{\rm DR}^2) -1 +\delta^{\rm FV}_3(X)\,,
\label{eq:app:tilde-l}
\end{eqnarray}
where $\mu_\text{DR}$ is the scale introduced by dimensional
regularization, and $X$ is squared mass (in physical units) of the
$\bar d d$ pion.
The functions $\delta_{1}^{FV}(X)$ and $\delta_{3}^{FV}(X)$ 
contain the finite volume corrections:
\begin{eqnarray}
\delta^{\rm FV}_1(M^2) &=& \frac4{ML} \sum_{n \ne 0}
\frac{K_1(|n| ML)}{|n|}
\label{eq:delta_1}
\\
\delta^{\rm FV}_3(M^2) &=& 2 \sum_{n\ne 0}
{K_0(|n| ML)}\,,
\label{eq:delta_3}
\end{eqnarray}
where $M$ is the pion mass, $L$ is the box size in the spatial
direction, $K_{1}$ and $K_{0}$ are modified Bessel functions,
and $n=(n_{1}, n_{2}, n_{3}, n_{4})$ is a image vector in
4-dimension lattice.
The norm $|n|$ is 
\begin{eqnarray}
|n| \equiv \sqrt{n_{1}^{2}+n_{2}^{2}+n_{3}^{2}
+\left(\frac{L_T}{L} n_4\right)^2}
\end{eqnarray}
where $L_T$ is the Euclidean temporal box size.\\
The details are explained in Ref.~\cite{wlee-10-3}.

\section{Numerical Study}
In order to calculate the finite volume corrections 
$\delta^{\rm FV}_1$ and $\delta^{\rm FV}_3$ in
Eqs.~(\ref{eq:delta_1}-\ref{eq:delta_3}), we use the following
criteria to truncate the sum over $n$.
For $\delta^{\rm FV}_1$, with desired precision 
$\epsilon = 1.0 \times 10^{-14}$ (double precision),
we first determine $r_{\tt{max}}$ from
\begin{equation}
[4\pi r_{\tt{max}}^2 ] \times \frac{K_1(r_{\tt{max}} ML)}{r_{\tt{max}}} 
= \epsilon  \times [6 K_1(ML)]\,.
\label{eq:r}
\end{equation}
Here, $4\pi r_{\tt{max}}^2$ is the density of image vectors 
at $|n| = r_{\tt{max}}$ and
$6 K_1(ML)$ is the contribution to $\delta^{\rm FV}_1$ 
from the first set of images with $|n| = 1$.
In words, we keep images out to a distance $r_{\tt{max}}$ at which the
contribution from a shell of radius $\Delta r=1$ equals the
desired precision times the leading contribution from $|n|=1$.
We assume in this estimate that $L_T\gg L$ (with $L_T$ the
extent in the temporal direction), so that we only consider
spatial images. This is the source of the factor of 6 multiplying
$K_1(ML)$.
Similarly for $\delta_{3}^{FV}$, we define $r_{\tt{max}}$
from 
\begin{equation}
[4\pi r_{\tt{max}}^2 ] \times K_0(r_{\tt{max}} ML) \ge 
\epsilon \times [6 K_0(ML)]\,.
\label{eq:r_0}
\end{equation}
In the second step, we define spatial and temporal ``radii''
through
\begin{equation}
r_{s}=r_{\tt{max}},
\qquad r_{t}=  \frac{L}{L_{T}} \times r_{\tt{max}} 
\,.
\end{equation}
Finally, when we calculate the finite volume corrections Eq.~\ref{eq:delta_1}
and Eq.~\ref{eq:delta_3}, we include only images
satisfying 
\begin{eqnarray}
&& -r_{s} \le n_{i} \le r_{s}  \qquad \text{for} \qquad 
i = 1,2,3 \nonumber\\
&& -r_{t} \le n_{4} \le r_{t}
\end{eqnarray}
Therefore, the number of the image vectors $n$ is essentially
$(2r_{s}+1)^{3} \times (2 r_{t}+1)$.

To draw plots of $B_{K}$ vs. pion mass-squared $X$, we 
calculate finite volume corrections for about hundred different 
mass values.
The radius $r_{\tt{max}}$ varies with $X$, but roughly
we find we need, for 100 different mass values in the relevant range,
about $10^9$ image vectors.
Since there are about 1000 configurations in each ensemble,
we need about $10^{12}$ evaluations of Bessel functions per ensemble.
If we use a standard CPU to calculate finite volume corrections for all
the MILC asqtad ensembles that we have data on, it takes about two months.
This is clearly impractical, and we need
significantly faster computational resources.
GPUs provide the solution to this problem.

\section{CUDA Programming}

GPUs are composed of many tiny multi-processors which can handle the
single instruction multiple data efficiently.
We use Nvidia GTX480 GPUs which have a peak speed of 168 giga flops
in double precision~\cite{wlee-10-5}.
We use CUDA for GPU programming and obtain 64.3 giga flops 
(38\% of the peak) in double precision.
This is almost 120 times faster than the CPU code (0.5 giga flops).
We use the following optimization techniques.
\begin{itemize}
\item Substituting Division by Multiplication: \\
Division is slower than multiplication in GPU calculation.
For example, the division operation $x / 4$ is much slower than the
multiplication operation $x \times 0.25$.
After this optimization, we get 16\% gain in the speed.

\item Coalesced Access:\\
Coalesced access allows sequential threads to access sequential 
GPU memories in parallel.
Coalesced access is at least twice as fast as uncoalesced access.
We find that including
coalesced access in the global sum algorithm leads to a 20\% gain.
\end{itemize}

\section{Results}
We use MILC asqtad ensembles listed in Table~\ref{tab:milc-lat}.
They are generated with $N_f = 2 + 1$ flavors of asqtad staggered
sea quarks.
\begin{table}[htb]
\caption{MILC lattices used for the numerical study.
Here, ``ens'' represents the number of gauge configurations,
``meas'' is the number of measurements per configuration,
and ID will be used later to identify the corresponding lattice.
\label{tab:milc-lat}}
\center
\begin{tabular}{l  l  l  c  l }
\hline
\hline
$a$ (fm) & $am_l/am_s$ & \ \ size & ens $\times$ meas  & ID \\
\hline
0.12 & 0.03/0.05  & $20^3 \times 64$ & $564 \times 9$  & C1 \\
0.12 & 0.02/0.05  & $20^3 \times 64$ & $486 \times 9$  & C2 \\
0.12 & 0.01/0.05  & $20^3 \times 64$ & $671 \times 9$  & C3 \\
0.12 & 0.01/0.05  & $28^3 \times 64$ & $275 \times 8$  & C3-2 \\
0.12 & 0.007/0.05 & $20^3 \times 64$ & $651 \times 10$ & C4 \\
0.12 & 0.005/0.05 & $24^3 \times 64$ & $509 \times 9$  & C5 \\
\hline
0.09 & 0.0062/0.031 & $28^3 \times 96$  & $995 \times 9$ & F1 \\
0.09 & 0.0031/0.031 & $40^3 \times 96$  & $850 \times 1$ & F2 \\
\hline
0.06 & 0.0036/0.018 & $48^3 \times 144$ & $744 \times 2$ & S1 \\
0.06 & 0.0025/0.018 & $56^3 \times 144$ & $198 \times 9$ & S2 \\
\hline
0.045 & 0.0028/0.014 & $64^3 \times 192$ & $705 \times 1$ & U1 \\
\hline
\hline
\end{tabular}
\end{table}
The values of light sea quark masses ($am_l$) and strange sea quark masses
($am_s$) are given in Table \ref{tab:milc-lat}.
We use four different lattice spacings: coarse ($a=0.12$ fm), fine
($a=0.09$ fm), superfine ($a=0.06$ fm), and ultrafine ($a=0.045$ fm)
lattices.

In our numerical study on $B_K$, we use HYP-smeared staggered fermions
as valence quarks.
HYP staggered fermions have a number of advantages such as reducing
taste symmetry breaking as efficiently as HISQ action
\cite{wlee-08-2}.
We use 10 different values of the valence quark masses
($m_x$ for the $d$ quark and $m_y$ for the $\bar s$) 
as given in Table \ref{tab:val-qmass}.
%
%
\begin{table}[htbp]
\caption{Valence quark masses (in lattice units).
\label{tab:val-qmass}}
\center
\begin{tabular}{ l  l  l}
\hline
\hline
$a$ (fm) & $a m_x$ and $a m_y$ & \\
\hline
0.12  &  $0.005 \times n$  & with $n=1,2,3,\ldots,10$ \\
0.09  &  $0.003 \times n$  & with $n=1,2,3,\ldots,10$ \\
0.06  &  $0.0018 \times n$ & with $n=1,2,3,\ldots,10$ \\
0.045 &  $0.0014 \times n$ & with $n=1,2,3,\ldots,10$ \\
\hline
\hline
\end{tabular}
\end{table}

In Table~\ref{tab:bk-fv}, we present our
results for $B_{K}$ with and without including finite volume (FV)
terms in the fitting, as well as the difference between the two.
Note that the differences are statistically significant despite
the fact that the error in the individual results is larger than
the difference. This is because the two fits are highly correlated.
We find very small shifts, indicating that FV effects are a subpercent
systematic. We also note that the impact of including FV corrections
on our largest lattice (C3-2) is negligible, indicating that this
volume is effectively infinite. 

\begin{table}[htbp]
\caption{ $B_K(\text{NDR},1/a)$ with finite volume corrections.
The results are obtained by extrapolation to physical down quark mass
and removing lattice artifacts due to taste breaking.
The second column gives the results from extrapolation using
the infinite volume SU(2) staggered ChPT form.
The third column gives results from fitting to the FV form,
and then removing the FV corrections from the final number.
The last column gives the percentage change.
The fit type is 4X3Y-NNLO of the SU(2) analysis, which is explained
in Ref.~\cite{wlee-10-3}.
$am_y$ is fixed to the heaviest quark mass (for example,
$am_y=0.05$ for the C3 ensemble).
\label{tab:bk-fv}}
\center
\begin{tabular}{ l l l l}
\hline
\hline
ID    & $B_K$       & $B_K$(FV)   & $\Delta B_K$ \\
\hline
C3    & 0.5734(46)  & 0.5743(46)  & +0.16\% \\
C3-2  & 0.5784(46)  & 0.5785(46)  & +0.02\% \\
F1    & 0.5074(37)  & 0.5049(37) & -0.49\% \\
S1    & 0.4914(65)  & 0.4898(65)  & -0.33\% \\
U1    & 0.4812(65)  & 0.4790(65)  & -0.46\% \\
\hline
\hline
\end{tabular}
\end{table}

\begin{figure}[!htbp]
\center
\subfigure[C3 ensemble]{\includegraphics[width=17pc]
{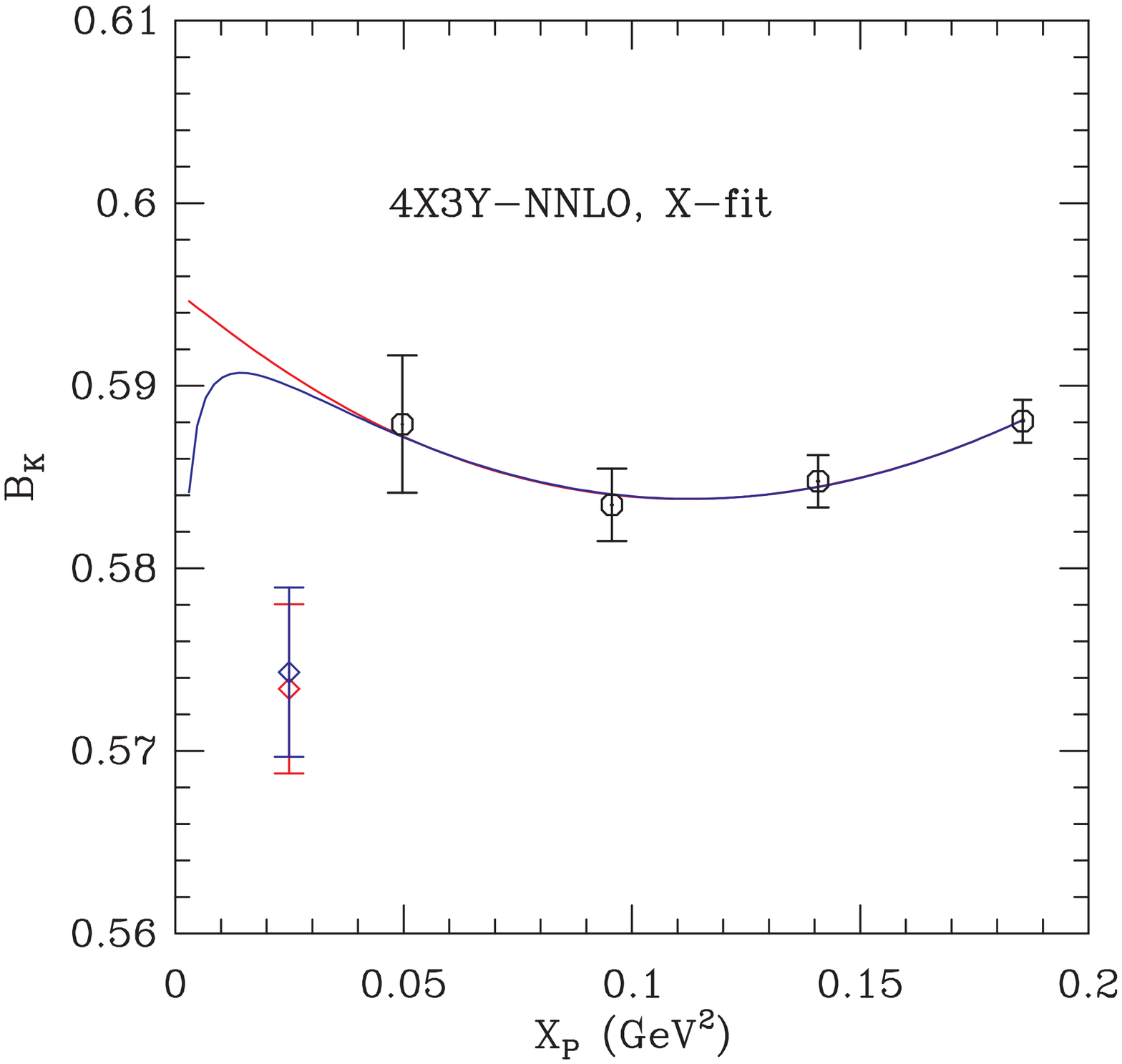}
\label{fig:C3}}
\subfigure[C3-2 ensemble]{\includegraphics[width=17pc]
{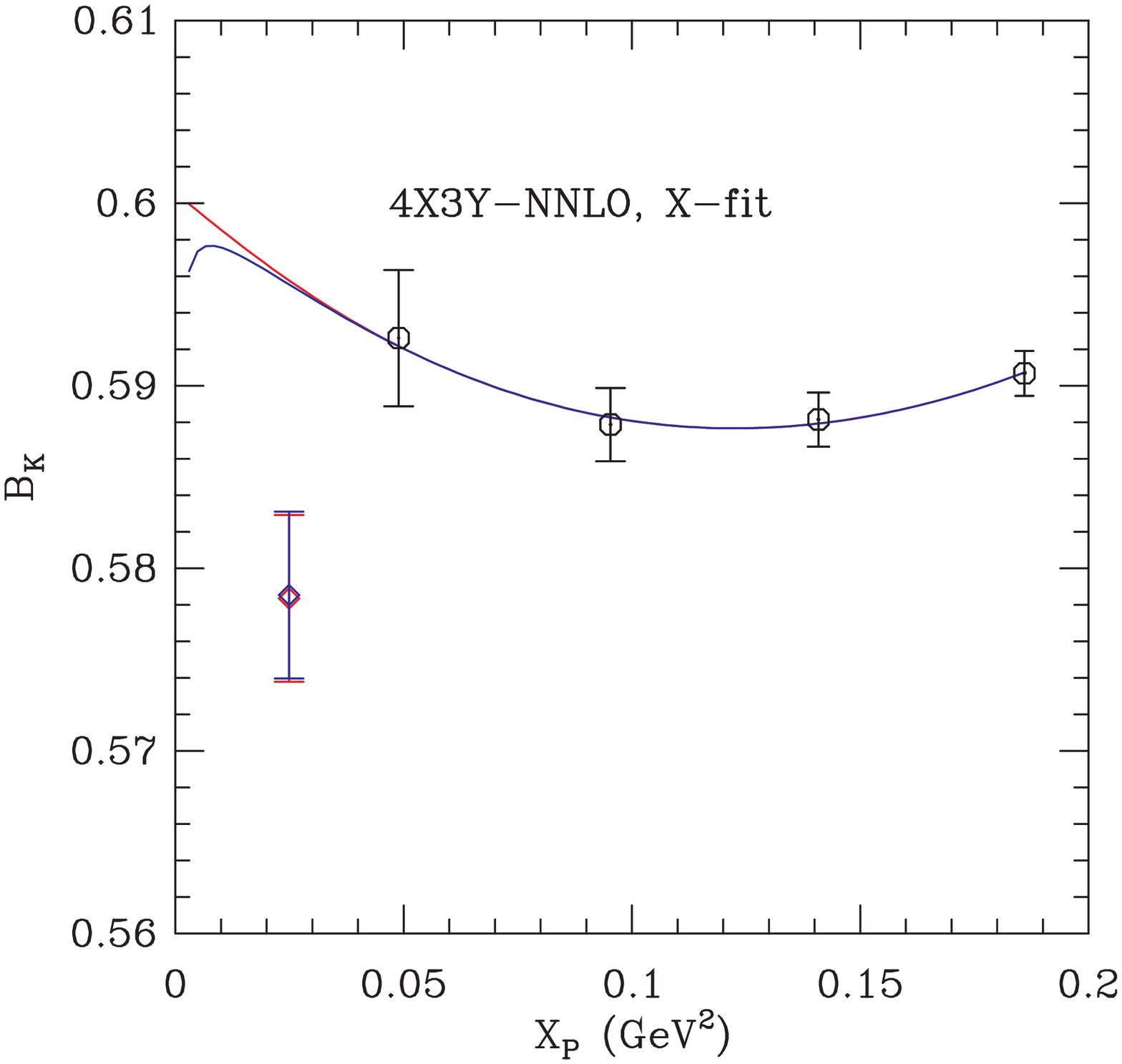}
\label{fig:C3-2}}

\caption{ $B_K(1/a)$ vs. $X$. The left figure shows results from the C3
  ensemble, while the right figure shows results from the C3-2 ensemble.  The
  fit type is 4X3Y-NNLO in the SU(2) analysis \cite{wlee-10-3}.  We
  fix $am_y = 0.05$.  The red line represents the results of fitting
  with no finite volume correction. The blue line corresponds to those
  with finite volume corrections included. The diamonds correspond to
  the $B_K$ value obtained by extrapolating $m_x$ to the physical
  light valence quark mass after setting all the pion taste-splittings 
  to zero. }
\end{figure}

\begin{figure}[!htbp]
\center

\subfigure[F1 ensemble]{\includegraphics[width=17pc]
{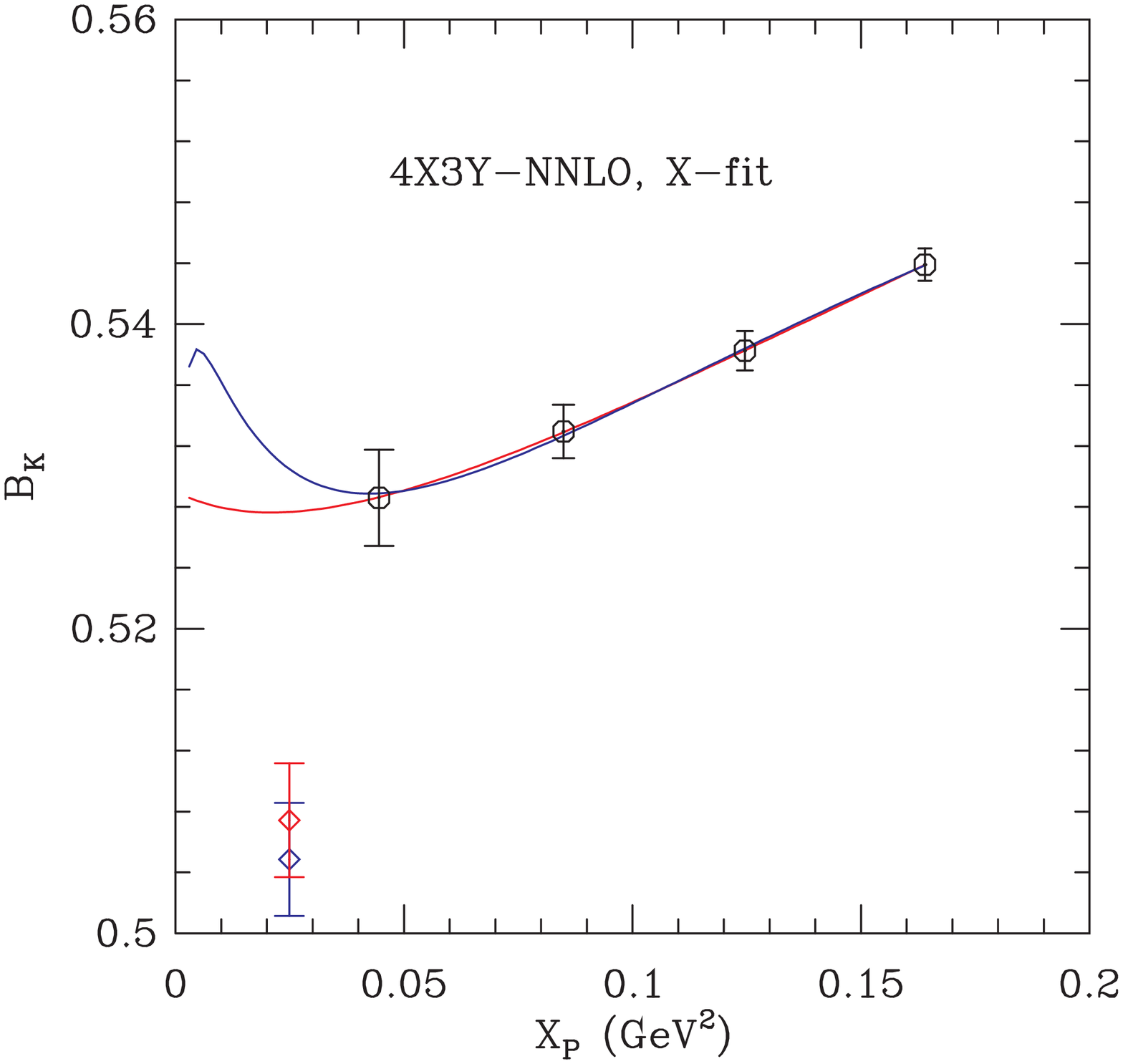}
\label{fig:F1}}
\subfigure[S1 ensemble]{\includegraphics[width=17pc]
{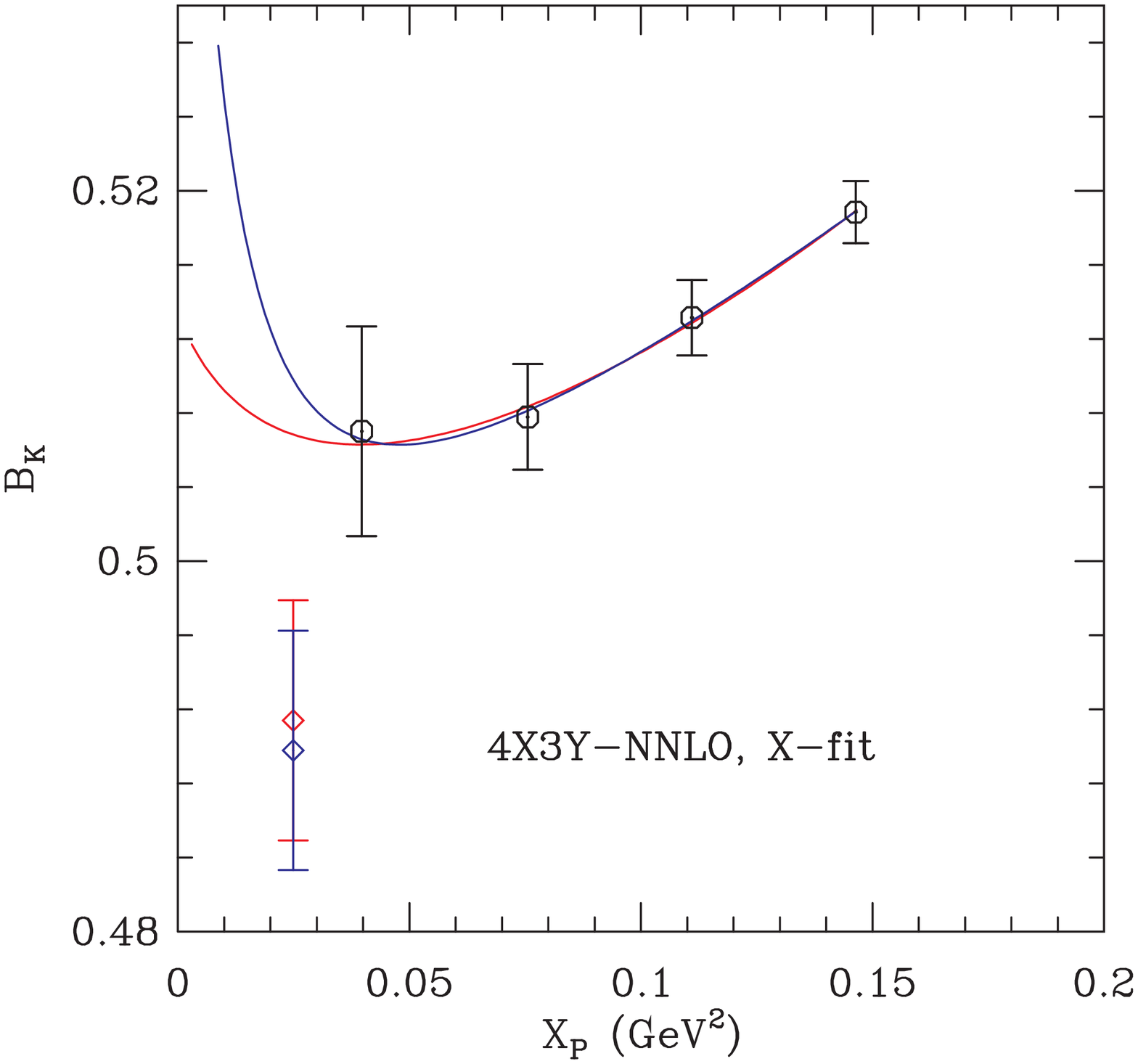}
\label{fig:S1}}

\caption{
$B_K(1/a)$ vs. $X$. The left figure shows results from the
F1 ensemble and the right figure from the S1 ensemble.
The fit type is 4X-NNLO in the SU(2) analysis \cite{wlee-10-3}.
}
\end{figure}

\begin{figure}[!htbp]
\center

\subfigure[U1 ensemble]{\includegraphics[width=17pc]
{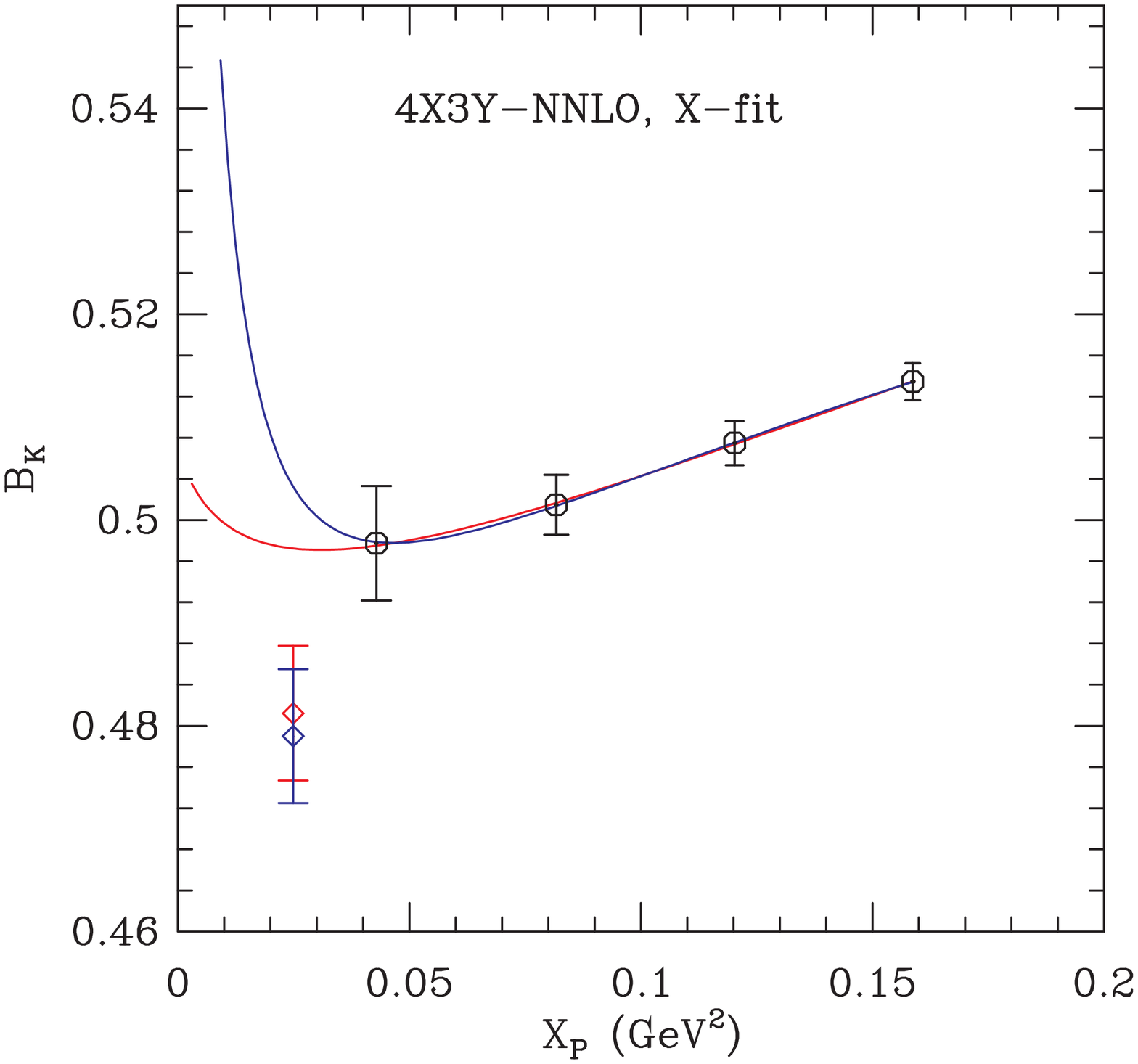}
\label{fig:U1}}
\subfigure[Scaling]{\includegraphics[width=17pc]
{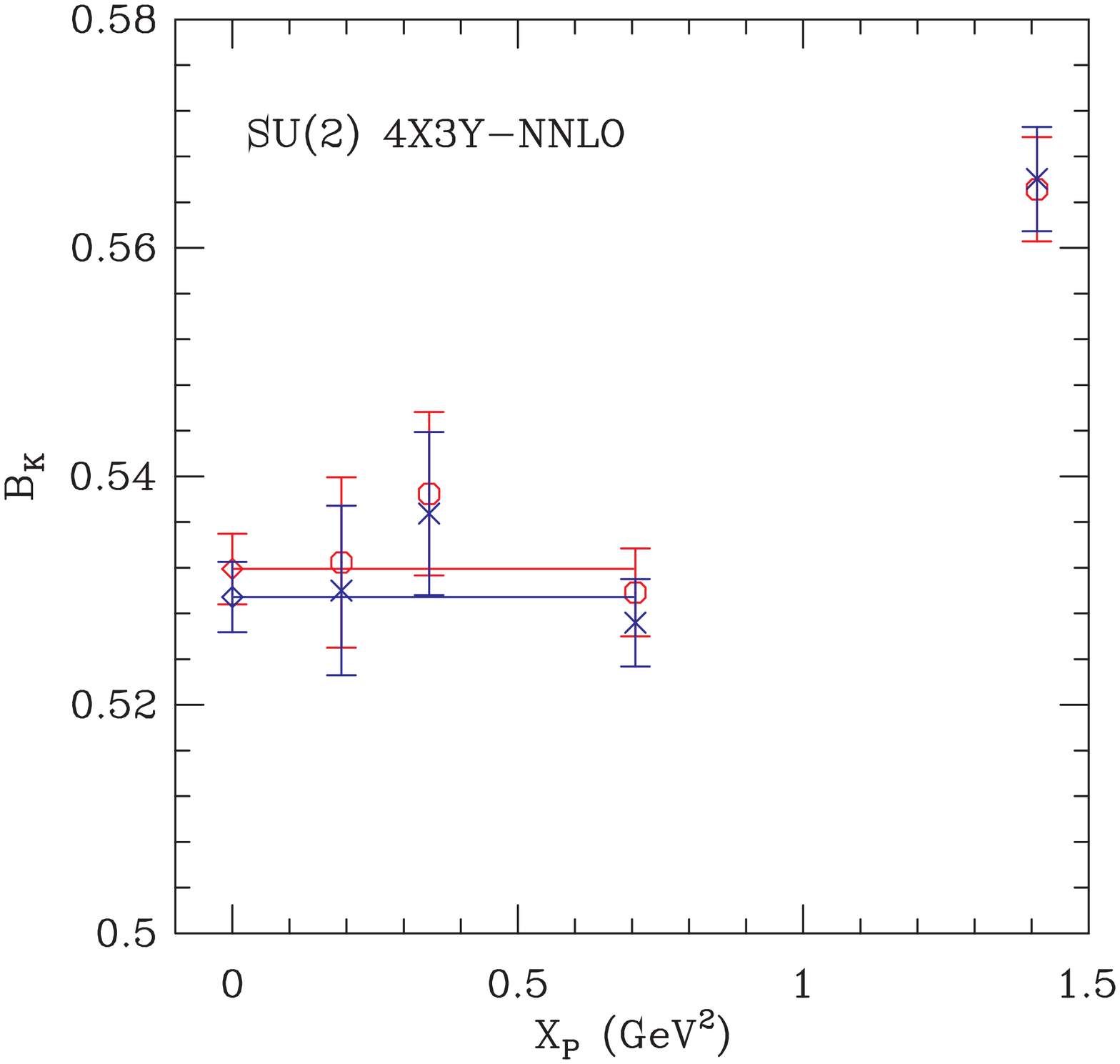}
\label{fig:scaling}}

\caption{
The left figure shows $B_K(1/a)$ vs. $X$ for the U1 ensemble. 
The fit type is 4X-NNLO in the SU(2) analysis. The right figure shows 
$B_K(2 {\rm GeV})$ vs. $a^2$. The red octagons show data 
obtained using the SU(2) fitting without the finite volume 
corrections.  The blue crosses show results from
SU(2) fitting with the FV corrections incorporated.  
Diamonds show the results after 
extrapolation to the continuum ($a=0$) using the smallest
three values of $a$.}
\end{figure}

We now display some of the fits that lead to these numbers.
Figures~\ref{fig:C3}, \ref{fig:C3-2}, 
\ref{fig:F1}, ~\ref{fig:S1} and ~\ref{fig:U1} 
show ``X-fits'' on the C3, C3-2, F1, S1 and U1 ensembles,
respectively. 
The red line denotes fitting without finite volume 
corrections and the blue line denotes those with FV corrections
included.
The diamonds give $B_{K}$ obtained, as explained above, by 
extrapolating $m_{x}\to m_d^{\rm phys}$,
setting all pion taste-splittings to zero, and
(in the case of the FV fit) setting $L,L_T \to \infty$.

Fig.~\ref{fig:scaling} compares the continuum extrapolation
with and without the finite volume corrections.
The total correction in the continuum limit is $0.46\%$. 

\section{Conclusion \label{sec:conclude}}

By using GPUs, we have significantly reduced the computational time for
FV corrections in NLO chiral expressions. This has made it
practical to fit and extrapolate using the FV-corrected forms.
Using this method we have updated all our SU(2) staggered ChPT fits,
with results reported in Ref.~\cite{wlee-11-2}.

We find the FV effect to be at the subpercent level, although, as
shown in the figures for the finest ensembles, FV effects would get much
larger if we lowered the valence quark masses any further.

Comparing the results from Table~\ref{tab:bk-fv} from the C3 and
C3-2 ensembles, we see that the FV shift on the C3 lattice
is significantly smaller than difference between the central
values from the two lattices. There is no inconsistency here
because the errors on individual lattices are large enough
that we cannot statistically distinguish between the results
on the two volumes. Because of this, we think that the FV
shift based on ChPT is a more reliable estimator of the FV
systematic, and we use this in our updated results.

\section{Acknowledgments}
C.~Jung is supported by the US DOE under contract DE-AC02-98CH10886.
The research of W.~Lee is supported by the Creative Research
Initiatives program (3348-20090015) of the NRF grant funded by the
Korean government (MEST).
W.~Lee would like to acknowledge the support from KISTI supercomputing
center through the strategic support program for the supercomputing
application research [No. KSC-2011-C3-03].
The work of S.~Sharpe is supported in part by the US DOE grant
no. DE-FG02-96ER40956.
Computations for this work were carried out in part on QCDOC computers 
of the USQCD Collaboration at Brookhaven National Laboratory.
The USQCD Collaboration are funded by the Office of
Science of the U.S. Department of Energy.


\begin{thebibliography}{99}
%
\bibitem{wlee-10-3}
Taegil Bae, \textit{et al.}, SWME Collaboration,
Phys. Rev. D\textbf{82}, (2010), 114509; 
[\texttt{arXiv:hep-lat/1008.5179}].  
%
\bibitem{wlee-11-1}
Jangho Kim, \textit{et al.}, SWME Collaboration,
Phys. Rev. D\textbf{83}, (2011), 117501; 
[\texttt{arXiv:hep-lat/1101.2685}].  
%
\bibitem{wlee-10-5}
Hyung-Jin Kim and Weonjong Lee, SWME Collaboration,
PoS (Lattice 2010) 230 ; [\texttt{arXiv:hep-lat/1010.4782}].  
%
\bibitem{wlee-08-2}
Taegil Bae and David H. Adams and Chulwoo Jung and
Hyung-Jin Kim and Jongjeong Kim and Kwangwoo Kim and
Weonjong Lee and Stephen R. Sharpe, SWME Collaboration,
Phys.~Rev.~D\textbf{77}, (2008), 094508 ; [\texttt{arXiv:0801.3000}].  
%
\bibitem{wlee-11-2}
Weonjong Lee, \textit{et al.}, SWME Collaboration,
PoS (Lattice 2011) 316 ; [\texttt{arXiv:hep-lat/1110.2576}].  
%
\end{thebibliography}
\end{document}